\documentclass[twocolumn]{jpsj2} 
\usepackage[dvips]{color}

\title{Variational Monte Carlo analysis of the Hubbard model with 
a confining potential: one-dimensional fermionic optical lattice systems}

\author{Yusuke \textsc{Fujihara},$^1$ Akihisa \textsc{Koga}$^1$ and 
Norio \textsc{Kawakami}$^{1,2}$}

\inst{$^1$Department of Applied Physics, Osaka University, 
Suita, Osaka 565-0871, Japan\\
$^2$Department of Physics, Kyoto University, Kyoto 606-8502, Japan
}

\abst{
We investigate the one-dimensional Hubbard model 
with a confining potential,
which may describe cold fermionic atoms trapped in an optical lattice.
Combining the variational Monte Carlo 
simulations with the new stochastic 
reconfiguration scheme proposed by Sorella, we present an efficient 
method to systematically treat the ground state properties of the 
confined system with a site-dependent potential.
By taking into account intersite correlations 
as well as site-dependent on-site correlations, 
we are able to describe the coexistence of the metallic and 
Mott insulating regions, which is consistent with 
other numerical results. Several possible improvements of the 
trial states are also addressed.
}

\kword{variational Monte Carlo, fermionic Hubbard model, 
optical lattice, ultracold atoms}

\begin{document}
\maketitle

\section{Introduction}

Strongly correlated particle systems have attracted much interest.
Among them, recent advances in laser cooling techniques make it possible 
to confine cold atoms in an artificial lattice, 
the so-called optical lattice. \cite{Greiner, Nature}
By tuning the amplitude and the direction of the laser, 
the hopping integral between sites and the strength of 
the confining potential 
can be controlled.
Furthermore, by means of the Feshbach resonance,\cite{Feshbach}
the interaction between particles is also controlled,
which thereby stimulates intensive experimental investigations 
on the effect of particle correlations in the optical lattice.

Bosonic systems have been discussed 
intensively\cite{Mandel, Stoferle2004, Paredes, Gerbier, Gerbier2006}
since the discovery of the superfluid-Mott insulator transition 
in the confined system with rubidium ions.\cite{Greiner}
The ground state properties of the system have theoretically been studied
in detail.\cite{Jaksch, NTTYamashita, Pollet, Bergkvist, Kovrizhin, Scarola, Wessel, Kollath}
Recently a degenerate Fermi gas with potassium ions (also lithium) is
also realized,
\cite{DeMarco, Truscott, Schreck, Granade, Hadzibabic, Greiner2004, Bartenstein, Zwierlein} which 
stimulates further investigations on strongly correlated 
fermionic systems in 
the optical lattice.\cite{Roati, Kohl, Stoferle}

Theoretical analyses of the fermionic systems 
have been done by means of the numerical techniques such as
the quantum Monte Carlo (QMC) method\cite{Rigol, RigolPRA, Pour} and
the density matrix renormalization group (DMRG).\cite{Yamashita, Yamashita2}
It has been elucidated that the metallic regions 
coexist with the Mott insulating regions 
in the one-dimensional (1D) confined system with large 
interactions.\cite{Rigol, RigolPRA, Yamashita}
However, these numerical methods face serious problems 
when they are applied to the 2D and 3D systems.
The DMRG method enables us to obtain the precise results only 
for the 1D systems, 
and the QMC method usually suffers from the minus 
sign problems for a large-cluster system.
Therefore, it is highly desirable to find a
powerful method to investigate
the fermionic confined systems systematically.

One of the potential methods suitable for this purpose may be
the variational Monte Carlo (VMC) calculation,
\cite{McMillan, Ceperley, Yokoyama, YokoyamaP}
which allows us to study not only 1D but also 2D and 3D 
systems in the same framework.
However, we again encounter some problems when the VMC method 
is extended to the confined systems with a site-dependent 
potential.
One of the most serious problems is that 
a large number of variational parameters should be treated 
in order to correctly describe the ground state properties.
Therefore, it is necessary to introduce sophisticated
techniques beyond the standard ones used so far in variational theory.

In this paper, we present an efficient method based on 
 the VMC simulations that is tractable and 
applicable to the optical lattice systems.
To overcome the above-mentioned difficulty inherent in the 
confined systems, we make use of a stochastic
reconfiguration with Hessian acceleration (SRH) scheme
to minimize the ground state energy in a given parameter space.
This method was recently proposed by Sorella \cite{Sorella} to
discuss long-range electron correlations in the periodic Hubbard system.
By taking the 1D Hubbard model with harmonic confinement as a
typical example, we will demonstrate that the new stochastic VMC simulations 
work quite well for the optical lattice system with a confining potential.

This paper is organized as follows.
In \S 2, we introduce the model Hamiltonian and 
briefly summarize the VMC method with the stochastic
reconfiguration scheme.
In \S 3, we apply the VMC simulations 
to discuss the effect of the particle correlations
in the confined system,
by using several distinct trial states.
A summary and discussions are given in the final section.

\section{Model and Method}

We consider a strongly correlated fermionic system in the optical lattice.
Here, we treat fermionic particles with $s=1/2$ spin, 
for simplicity, although a potassium atom $^{40}$K, for example,  
has an $s=9/2$ spin.\cite{Kohl}
In this paper, we restrict our discussions to the 1D 
 model in order to clearly demonstrate the potentiality of our 
numerical method applicable to the confined systems with a
site-dependent potential.
The simplified model Hamiltonian we consider here reads
\cite{Rigol, RigolPRA}
\begin{equation}
\mathcal{H} = -t \sum_{<i,j>\sigma} c^{\dagger}_{i \sigma} c_{j \sigma}
 + \sum_{i \sigma} V_{i} n_{i \sigma}
 + U \sum_{i} n_{i \uparrow} n_{i \downarrow},
\label{Hamiltonian}
\end{equation}
where $c^{\dagger}_{i \sigma}$ ($c_{i \sigma}$) is a creation (annihilation) 
operator at the $i$th site with spin $\sigma (= \uparrow, \downarrow)$, 
and $n_{i \sigma} = c^{\dagger}_{i \sigma} c_{i \sigma}$.
$t$ is the nearest neighbor hopping matrix, and
$U(>0)$ the on-site repulsive interaction. 
Interacting fermions are assumed to be trapped by 
the harmonic potential $V_i$, where 
$V_{L/2-1} = V_{L/2}=0$ at two sites around the center, 
$V_{0} = V_{L-1}= V$ at the edges ($L$ is the length of the system).
In the following, we fix $t=1$ as an energy unit.

In the confined system,
the local particle count depends on the site because the confining potential
$V_i$ is site-dependent.
Namely, the particle density around the center 
is larger than that around the edge. 
This suggests that the introduction of 
the repulsive interaction changes the distribution of 
various local quantities such as the particle density 
and the double occupancy.
Therefore, the careful treatment of the particle
correlations is desirable to clarify the ground state properties.

To this end,
we make use of the VMC method.
\cite{McMillan, Ceperley, Yokoyama, YokoyamaP}
In comparison with other numerical methods,
it has an advantage in extending the calculation to 
the 2D and 3D fermionic systems, as mentioned above.
This method has successfully been applied to 
various correlated electron systems with periodic potentials.
\cite{Giamarchi,Oguri,Otsuka,Liu,WatanabeII,Kuratani,Kobayashi,Koga,WatanabeI,McQueen,Kono}

We introduce here 
a trial state with Jastrow-type projection operators as\cite{Jastrow}
\begin{equation}
| \Psi \rangle
 = \mathcal{J}
| \Phi \rangle,
\label{WF}
\end{equation}
where $| \Phi \rangle$ is the Slater determinant for the non-interacting
state and the Jastrow operator $\mathcal{J}$
 incorporates two-particle correlations,
whose explicit form will be given in the next section.
In order to include the effects of the inhomogeneous 
potential, we should deal with a large number of site-dependent 
variational parameters. This is contrasted to the 
standard problems in condensed matter physics, 
\cite{McMillan,Ceperley,Yokoyama,YokoyamaP,Sorella,Giamarchi,
Oguri,Otsuka,Liu,WatanabeII,Kuratani,Kobayashi,Koga,WatanabeI,McQueen,Kono}
where only one or two parameters are sufficient to describe the
ground state properties. 
 Although those parameters are invaluable to obtain a more accurate solution
in the confined systems,
it becomes much more difficult to determine their optimized values.

For optimizing the parameters, there exist some iterative methods.
In ordinary iterative methods, the set of parameters $\mathbf{x}$ can be iteratively changed by $\mathbf{x} \rightarrow \mathbf{x} + \boldsymbol{\gamma}$.
Here $\boldsymbol{\gamma}$ is determined by minimizing the energy
\begin{eqnarray}
\Delta E &=& E(\mathbf{x}+\boldsymbol{\gamma}) - E(\mathbf{x}) \nonumber \\
         &=& \frac{1}{2} \boldsymbol{\gamma}^{T} B \boldsymbol{\gamma} - \mathbf{f}^T \boldsymbol{\gamma}, \label{of}
\end{eqnarray}
where $\mathbf{f}$ and $B$ denote the first energy derivative
 and the Hessian matrix.
As a result, $\boldsymbol{\gamma}$ is denoted as
\begin{equation}
\boldsymbol{\gamma} = B^{-1} \mathbf{f}.
\label{gamma}
\end{equation}

In order to optimize the parameters efficiently, it is important 
to evaluate $\mathbf{f}$ and $B$.
The SRH scheme, which is based on the iterative methods, 
 allows us to optimize enormous parameters even more
 efficiently.\cite{Sorella}
In comparison with other iterative techniques
such as the Powell's method and the quasi-Newton method,
it has an advantage in reaching the ground state in a few iterations.
The important idea in this scheme is that $\mathbf{f}$ and $B$ are evaluated by using the statistical fluctuations of each projection operator, and in the framework of VMC, the efficient and rapid convergence is achieved.
Furthermore, we can introduce an additional free  parameter $\beta$ in 
the matrix $B$, which can 
accelerate the convergence. This parameter originates from the expansion 
coefficient of the trial wave function, and contributes to the efficiency and 
the accuracy of the calculation.
This scheme was originally introduced to analyze a trial wave function with 
longer-range density-density or spin-spin correlations in the 
periodic systems.\cite{Sorella}

In this paper, we extend the SRH scheme so as to treat 
 multivariable trial wave functions including
 the effects of the inhomogeneous potential.
When the site-dependent parameters are introduced, the Hessian matrix $B$ 
should become positive semidefinite with some zero eigenvalues.
This reflects the fact that no particle exists around edges 
in the confining potential.
In this case, some parameters do not contribute to the energy and thus
cannot be determined straightforwardly.
Therefore, it may be difficult to obtain $B^{-1}$ in eq. (\ref{gamma}).
To overcome this difficulty, we propose to make use of
\begin{equation}
\boldsymbol{\gamma} = B^{+} \mathbf{f}
\label{gamma2}
\end{equation}
instead of eq. (\ref{gamma}).
Here $B^{+}$ is the Moore-Penrose pseudo-inverse matrix of $B$.
This matrix is always defined and has similar properties to 
the inverse matrix.  We will demonstrate below that
this new stochastic scheme works well and thereby enables us to discuss
 how particle correlations affect the ground state properties 
in the confined system. 

\section{Results}

In this section, we perform the VMC simulations with several trial states 
to clarify the role of the on-site and intersite particle correlations, 
by which we confirm the potentiality of our method.
Here, we deal with the system ($L=100$)
with $N_{\uparrow} = N_{\downarrow} = 30$ and $V = 10$,
where $N_\sigma$ is the number of particles with spin $\sigma$.

\subsection{Gutzwiller paramagnetic state}

First, we consider the Gutzwiller paramagnetic trial state,
which incorporates the on-site particle correlations.
The Gutzwiller trial state\cite{Gutzwiller, Kaplan, Fazekas} 
is explicitly given as
\begin{eqnarray}
| \Psi_{G} \rangle&=& \mathcal{P}_{G} | \Phi_{0} \rangle,\label{gwf} \\
\mathcal{P}_{G} &=&
\exp \left[ \sum_{i} \alpha_{i} n_{i\uparrow} n_{i\downarrow} \right],
\label{PG}
\end{eqnarray}
where $| \Phi_{0} \rangle$ is the Slater determinant 
for the noninteracting paramagnetic ground state.
Note that 
$\alpha_{i} (i=0, \ldots, L-1)$ 
is a site-dependent variational parameter with $\alpha_i=\alpha_{L-i-1}$,
which  may 
describe the effect of local correlations in our inhomogeneous system.

By optimizing $L/2(=50)$ variational parameters $\{\alpha_i\}$  
by means of the SRH scheme, 
we calculate local quantities for the confined system.
Let us first look at how efficiently our scheme works for optimizing 
the variational parameters.
\begin{figure}[tb]
\begin{center}
\includegraphics[width=7cm]{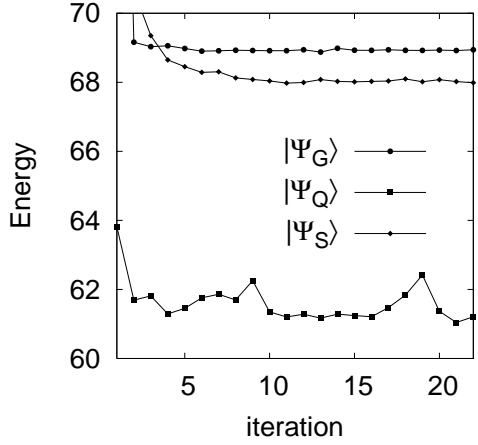}
\end{center}
\caption{Calculated energies in each iteration.
Filled circles, squares and diamonds are the results obtained
 from three distinct trial states (see text)
  when $U=6$ and $V=10$.}
\label{ERROR}
\end{figure}
In Fig. \ref{ERROR},
we show the calculated energy as a function of the iteration
when it starts from a certain initial state
($| \Psi_{Q} \rangle$ and $| \Psi_{S} \rangle$ are defined 
in the following subsections).
For each iteration, we perform the VMC calculation with large samplings 
$(\sim 500,000)$.
An important point is that the ground state  almost converges
in a few iterations, although 
the energy for each trial state slightly oscillates 
even after sufficient iterations.
We find that this does not depend on the number of 
the Monte Carlo samplings for each iteration, which 
implies that statistical errors in our calculations mainly come from 
this oscillation.  Anyway, we find that the stochastic
VMC scheme with eq. (\ref{gamma2}) works quite well for our
confined system.

\begin{figure}[htb]
\begin{tabular}{cc}
\hspace{-5mm}
\resizebox{46mm}{!}{\includegraphics{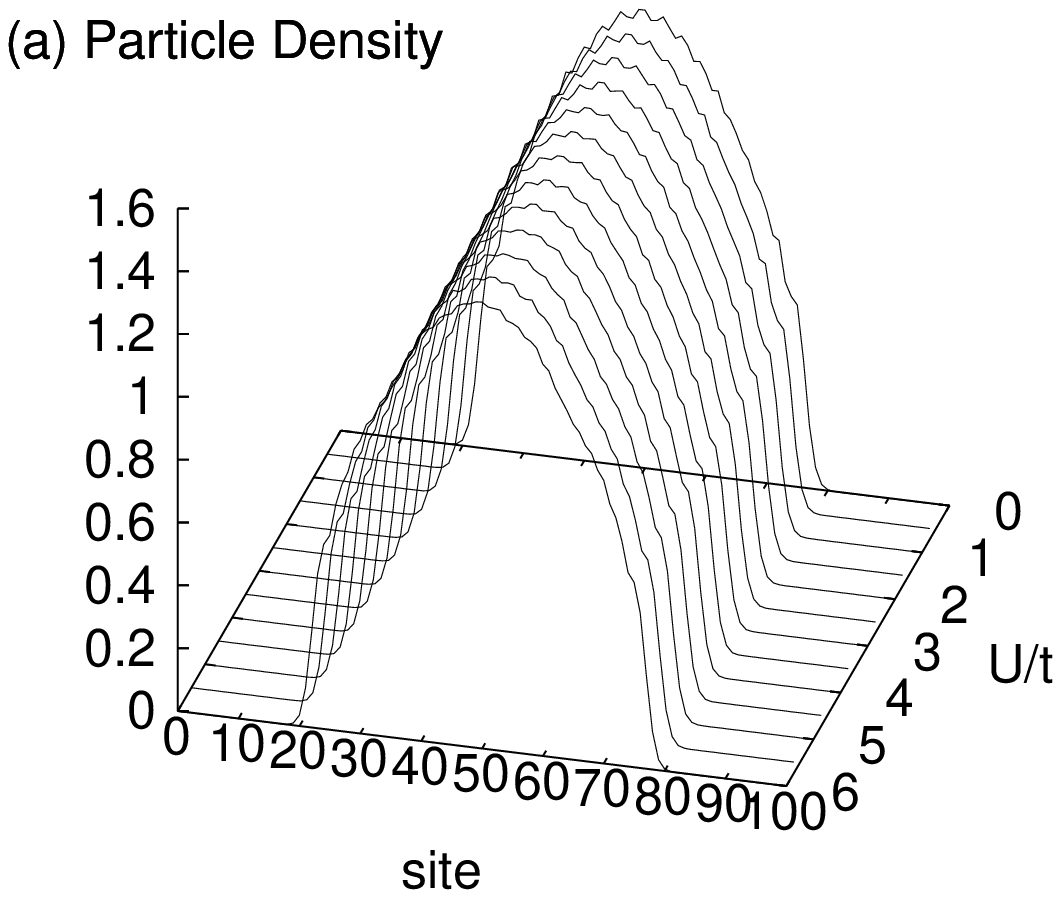}} &
\hspace{-7mm}
\resizebox{46mm}{!}{\includegraphics{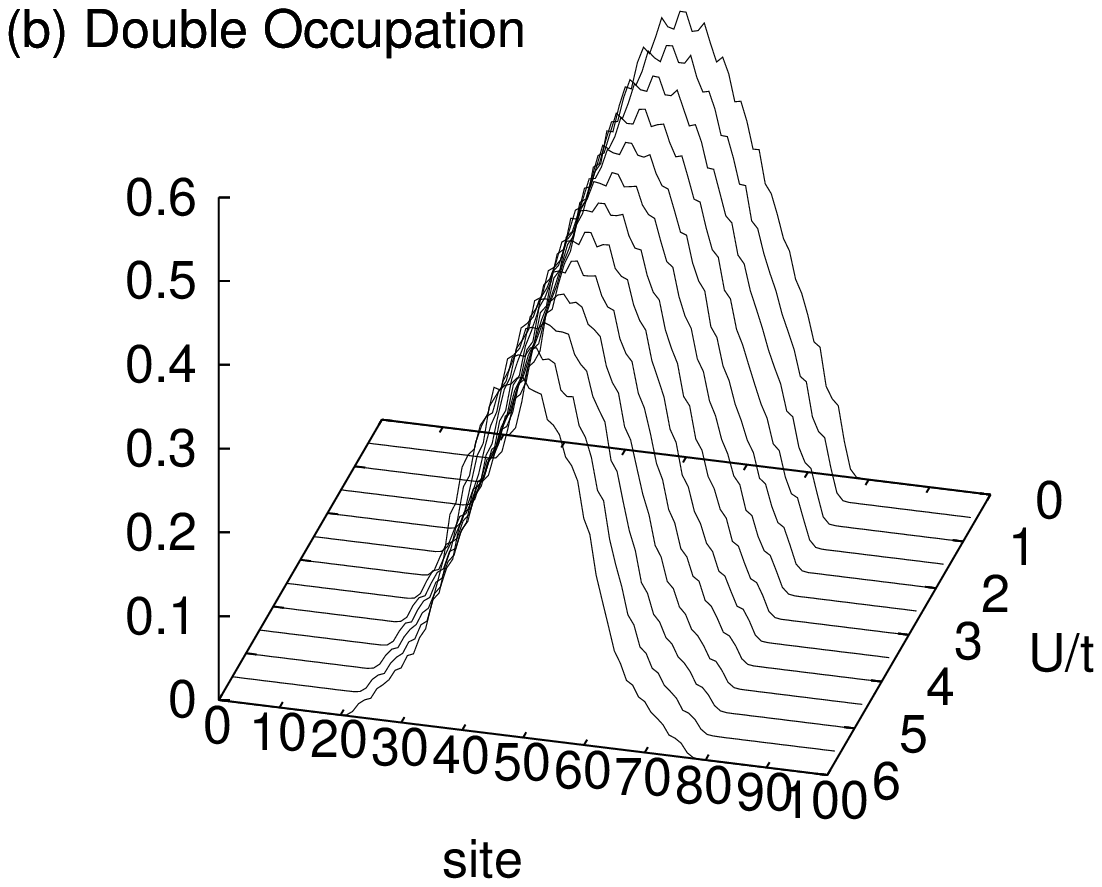}} \\
\hspace{-5mm}
\resizebox{46mm}{!}{\includegraphics{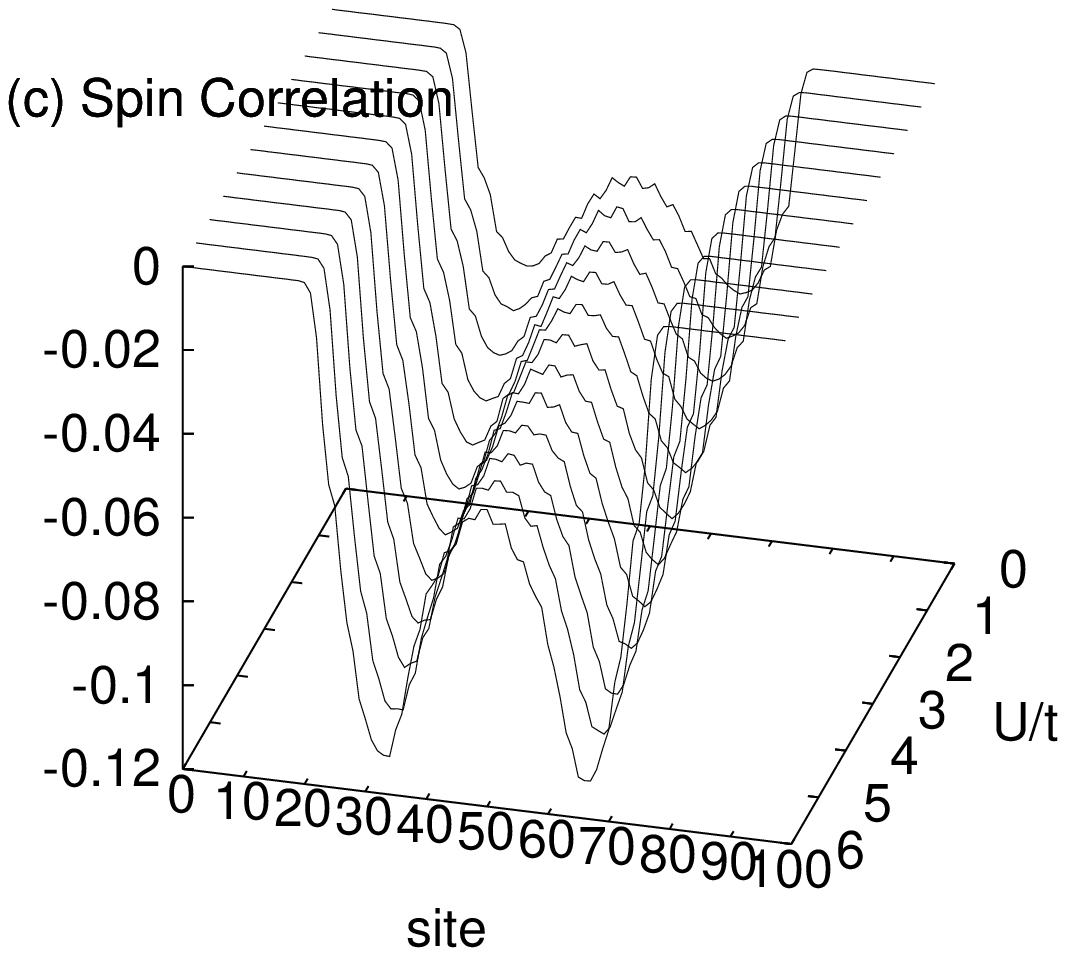}} &
\hspace{-7mm}
\resizebox{46mm}{!}{\includegraphics{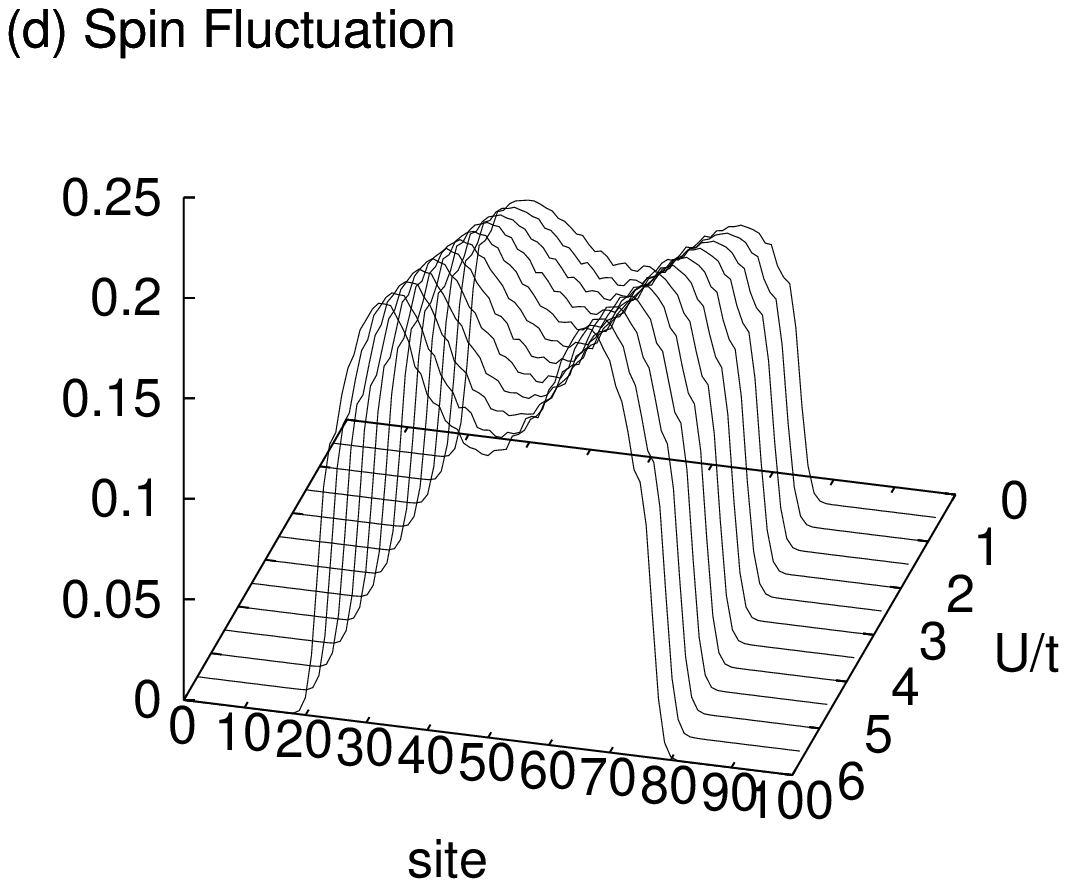}} \\
\hspace{-5mm}
\resizebox{46mm}{!}{\includegraphics{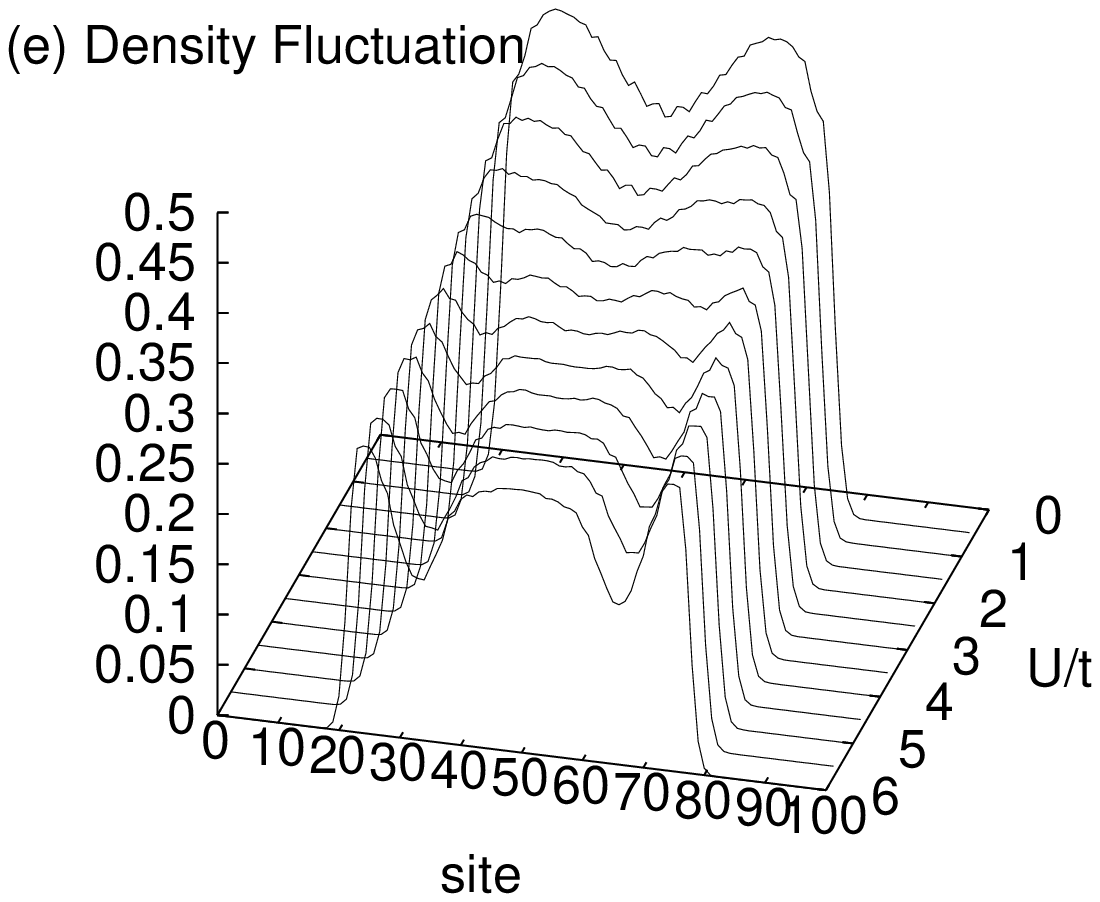}} \\
\end{tabular}
\caption{(a) particle density, (b) double occupation, 
(c) short-range spin correlation, (d) spin fluctuations, and 
(e) density fluctuations, which are obtained by the 
Gutzwiller wave function $| \Psi_{G} \rangle$.}
\label{GPQ}
\end{figure}

Several physical quantities thus calculated are summarized 
 in Fig. \ref{GPQ}.
It is found that
the introduction of the interaction slightly changes 
the local particle density $\langle n_i\rangle$, and 
decreases the local double occupation 
$\langle n_{i\uparrow}n_{i\downarrow}\rangle$.
At the same time,
spin fluctuations $\langle (S_i^z-\langle S_i^z\rangle )^2\rangle$ 
are enhanced and 
density fluctuations $\langle (n_i-\langle n_i\rangle)^2\rangle$ 
are suppressed.
This implies that the repulsive interaction $U$ 
excludes
doubly occupied states at each lattice site.
It should be noticed that nearest-neighbor spin correlations
$\langle (S^{z}_{i}S^{z}_{i-1}+S^{z}_{i}S^{z}_{i-1})/2 \rangle$
are strongly enhanced in the two regions around the 30th and 70th sites.
In those regions, the average of the particle 
density is almost unity and 
density (spin) fluctuations are suppressed (enhanced) strongly.
Namely, the repulsive interaction $U$ 
has a tendency to localize particles and 
enhance local spin fluctuations, as should be expected.

\begin{figure}[htb]
\begin{center}
\includegraphics[width=9cm]{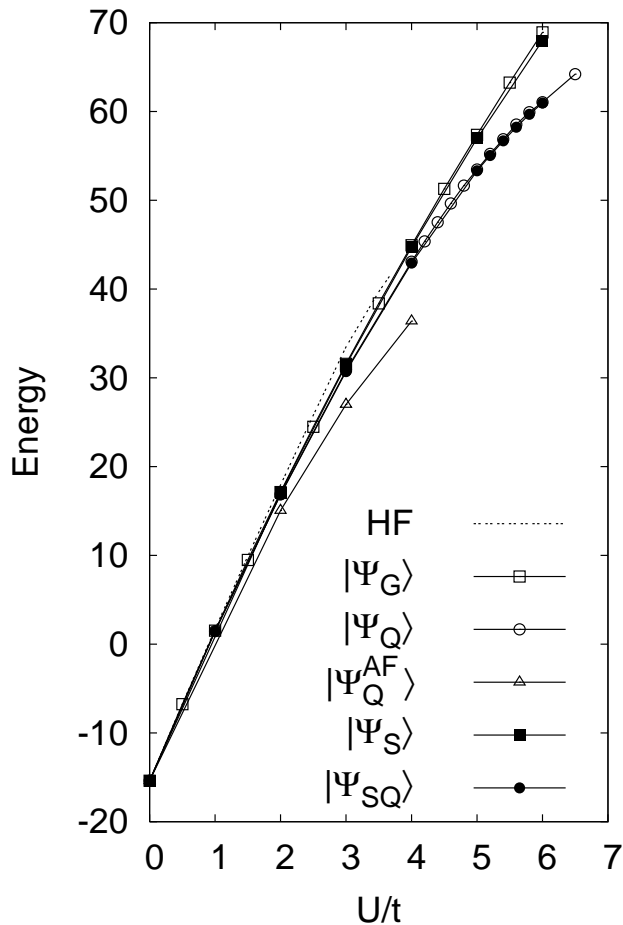}
\end{center}
\caption{The ground state energy
as a function of the repulsive interaction $U$.
Open squares, open circles, open triangles, filled squares, and filled circles 
are the results obtained from the Gutzwiller(GW), paramagnetic DH+GW,
magnetically ordered DH+GW, SC+GW and SC+DH+GW trial state (see text for
each notation).
The energy obtained by the Hartree-Fock (HF) approximation is
shown by the dotted line, for comparison. 
Statistical errors originating from stochastic oscillations
 are less than the size of symbol for each point.
}
\label{Egraph}
\end{figure}

In this way, the trial state with site-dependent Gutzwiller 
factors can take into account the local particle correlations to 
some extent. In fact, the ground state energy obtained 
from the Gutzwiller trial state eq. (\ref{gwf})
is lower than that obtained by the ordinary Hartree-Fock approximation,
as shown in Fig. \ref{Egraph}.
However, we could not reproduce 
the Mott insulating region that should be spatially extended 
in a {\it finite} region, 	
as suggested by other groups.\cite{Rigol, RigolPRA, Yamashita}

\subsection{Intersite doublon-holon correlations}
To take into account the particle correlations more precisely,
we introduce the intersite doublon-holon (DH) correlations.
\cite{Yokoyama, YokoyamaP}
The corresponding trial state is given as,
\begin{eqnarray}
| \Psi_{Q} \rangle &=& \mathcal{P}_{Q} \mathcal{P}_{G} | \Phi_{0} \rangle, 
\label{sDHwf}\\
\mathcal{P}_{Q} &=&
\exp \left[ \alpha' \sum_{i} \widehat{Q}_{i} \right],
\label{PQ}
\end{eqnarray}
where
\begin{eqnarray}
\widehat{Q}_{i} = 
	\widehat{D}_{i} \displaystyle\prod_{\tau} (1-\widehat{H}_{i+\tau})
	 +\widehat{H}_{i} \displaystyle\prod_{\tau}
	 (1-\widehat{D}_{i+\tau}) 
\label{Q}
\end{eqnarray}
with
$\widehat{D}_{i} = n_{i \uparrow} n_{i \downarrow}$,
$\widehat{H}_{i} = (1-n_{i \uparrow})(1-n_{i \downarrow})$
and $\tau$ runs over all the nearest neighbors.
The variational parameter $\alpha'$ is assumed to be
independent on the sites, for simplicity (see the discussion in \ref{3.3}).
Note that the doublon-holon projection operator 
$\mathcal{P}_{Q}$  takes into account the nearest-neighbor
correlations between a doubly occupied site and an empty site,
which may be important to stabilize the Mott insulating state.

\begin{figure}[htb]
\begin{tabular}{cc}
\hspace{-5mm}
\resizebox{46mm}{!}{\includegraphics{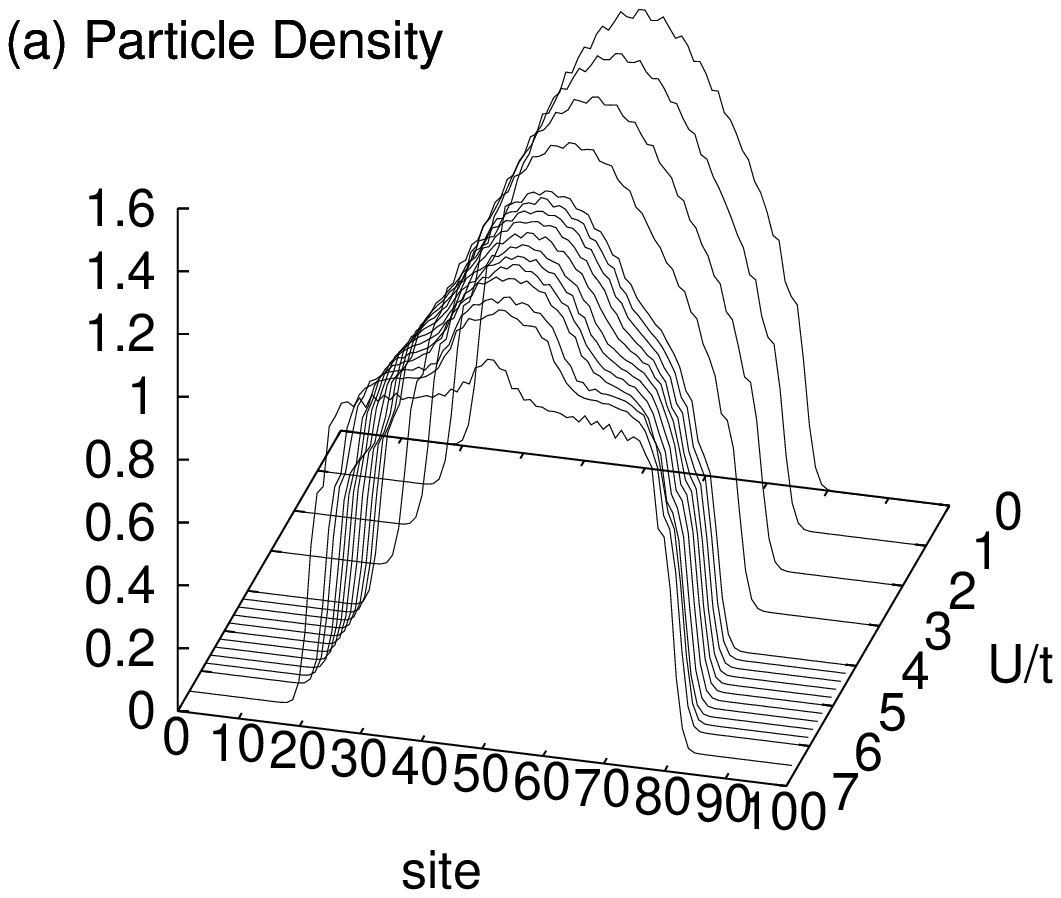}} &
\hspace{-7mm}
\resizebox{46mm}{!}{\includegraphics{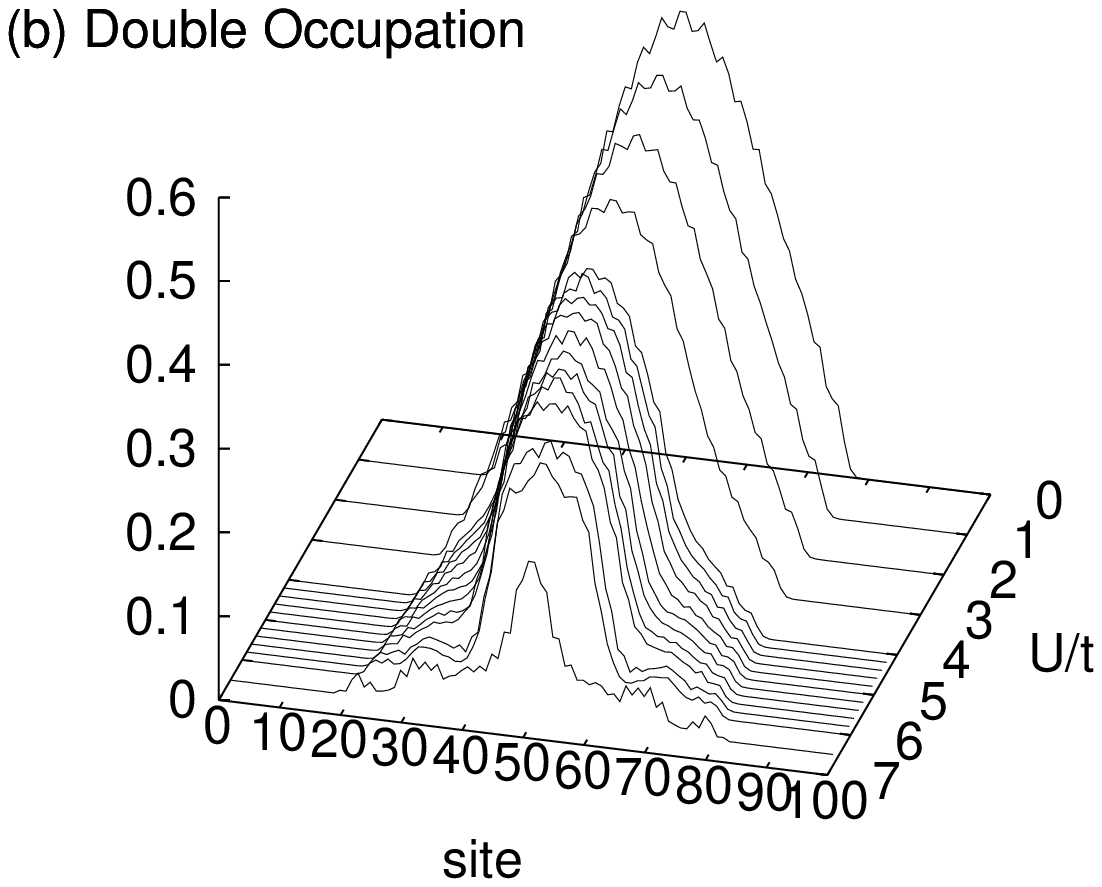}} \\
\hspace{-5mm}
\resizebox{46mm}{!}{\includegraphics{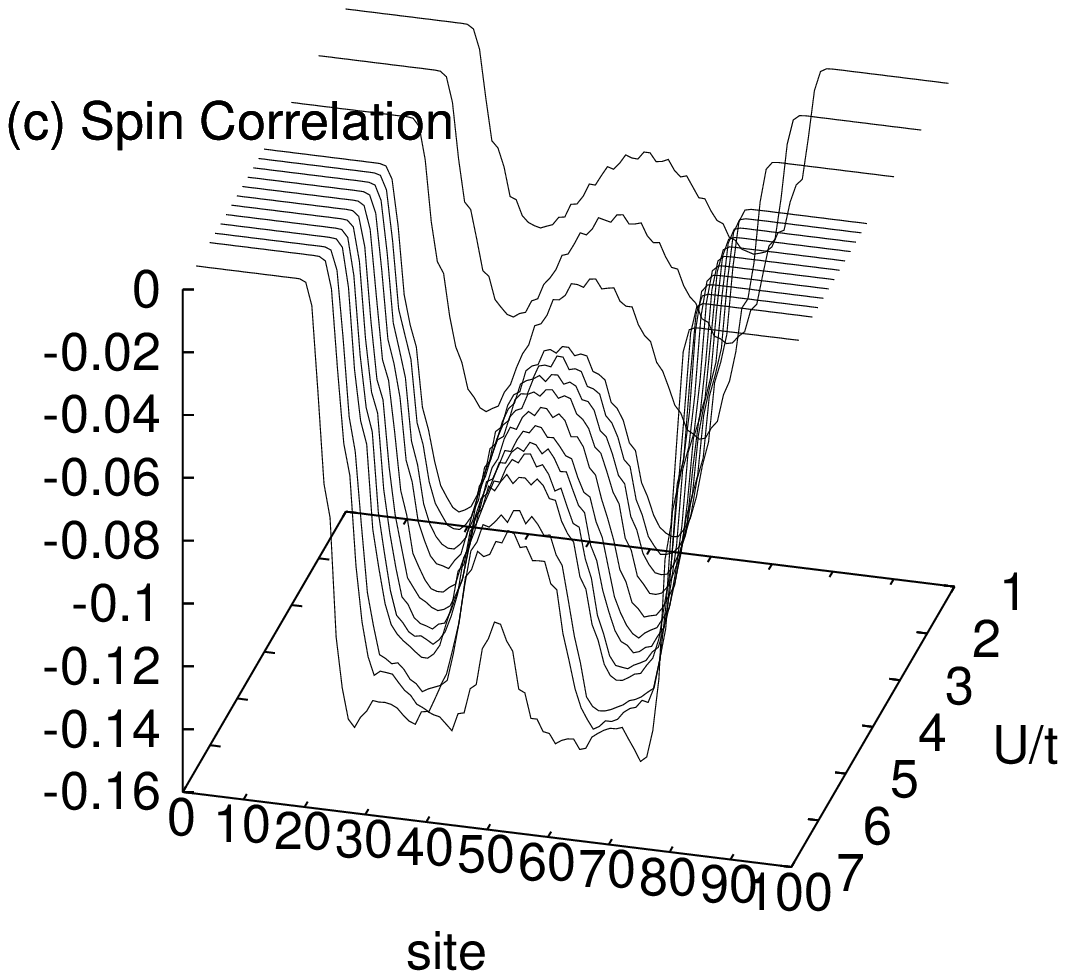}} &
\hspace{-7mm}
\resizebox{46mm}{!}{\includegraphics{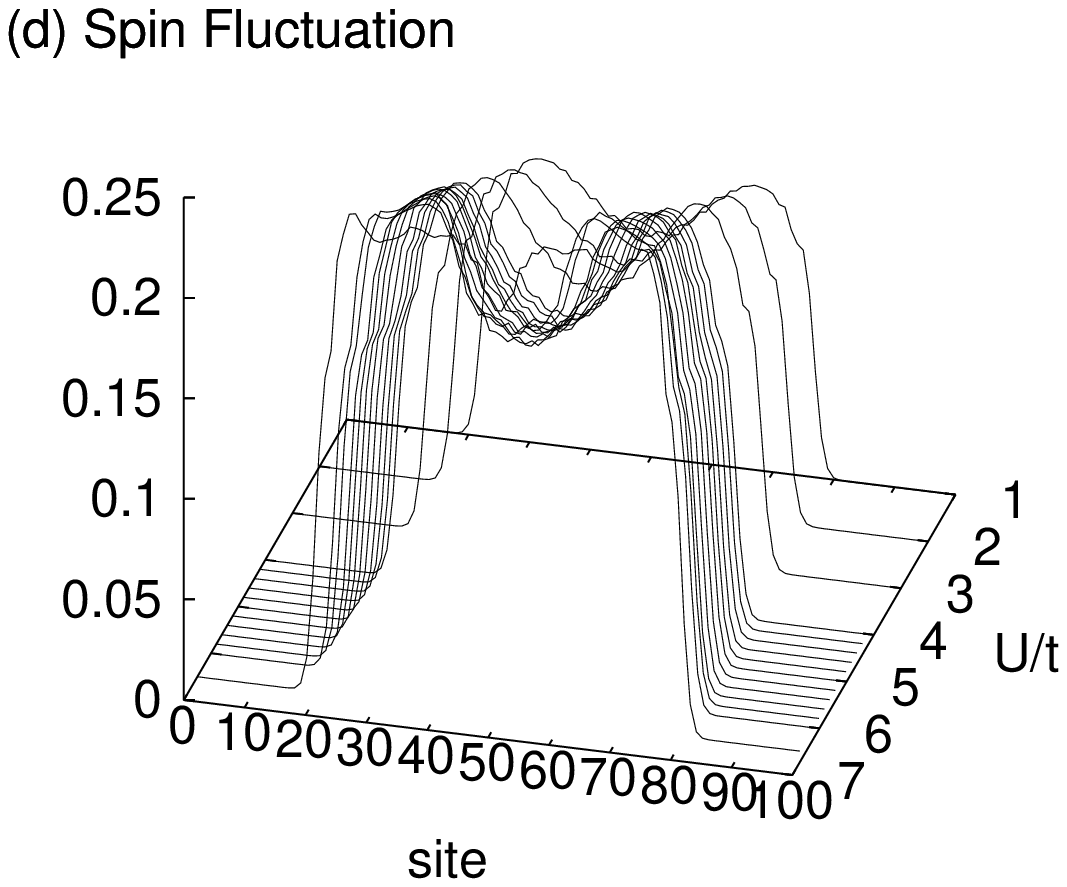}} \\
\hspace{-5mm}
\resizebox{46mm}{!}{\includegraphics{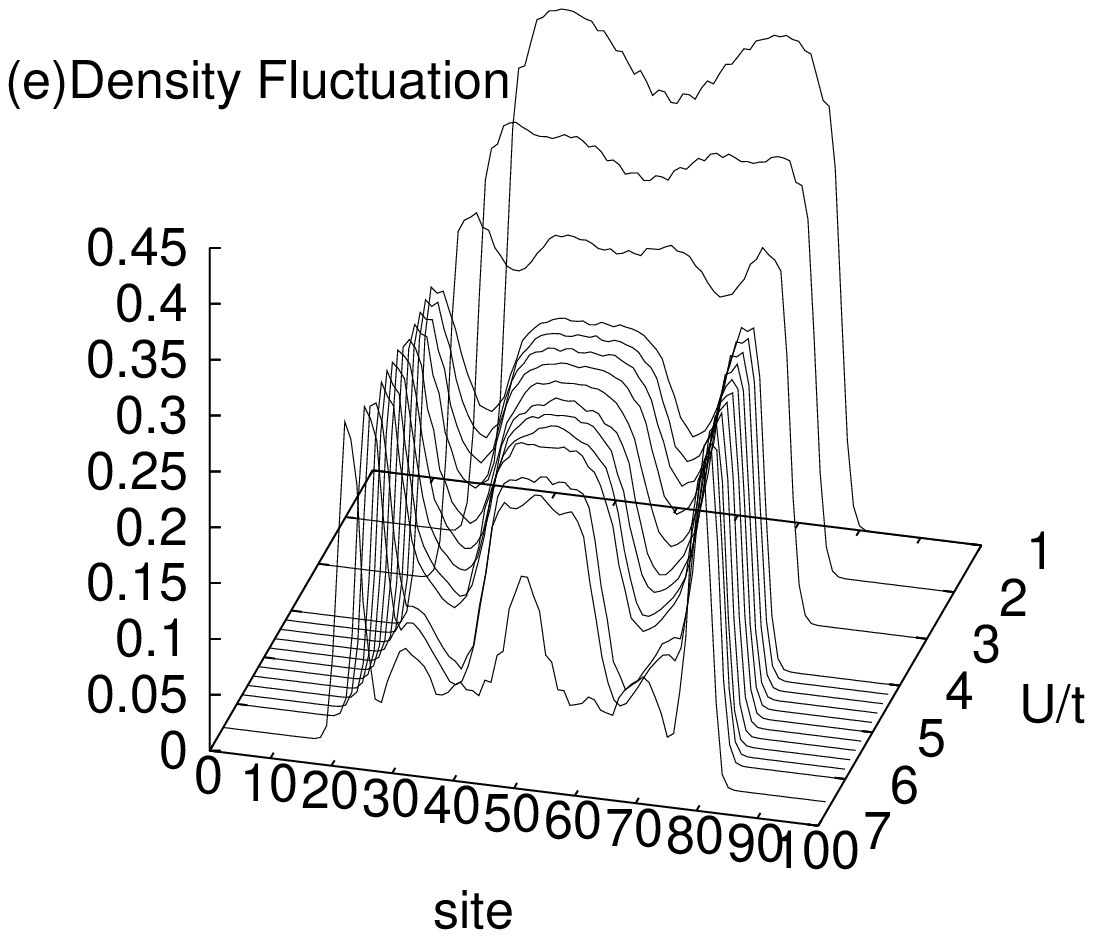}} \\
\end{tabular}
\caption{(a) particle density, (b) double occupation, 
(c) short range spin correlation, 
(d) spin fluctuations, and (e) density fluctuations,
which are obtained from the DH+GW trial state, $| \Psi_{Q} \rangle$.}
\label{sDHPQ}
\end{figure}

The ground state energy thus computed is shown in Fig. \ref{Egraph}.
It is found that the energies obtained from the trial states 
eqs. (\ref{gwf}) and (\ref{sDHwf}) are almost the same 
in the weak coupling region.
On the other hand, in the strong coupling region $U>4$,
we find that the ground state energy for the trial 
state (\ref{sDHwf}) is 
lower than the other, implying that the DH 
correlations play an important role in the region. 
In fact, characteristic behavior appears 
in various physical quantities beyond $U=4$, 
as shown in Fig. \ref{sDHPQ}.
It is seen that when the interaction is increased, 
the plateau regions appear in the curve of the particle density 
at $\langle n_{i} \rangle =1$, where the double occupancy is suppressed 
strongly.
Furthermore, we find that spin (density) fluctuations are considerably
enhanced (suppressed).
This implies that the commensurability (half-filling 
condition) is induced self-consistently
due to particle correlations and thus the 
Mott insulating states are stabilized in {\it finite} regions
in contrast to the results of the Gutzwiller trial 
state (see Fig. \ref{GPQ}).

\begin{figure}[htb]
\begin{center}
\includegraphics[width=8cm]{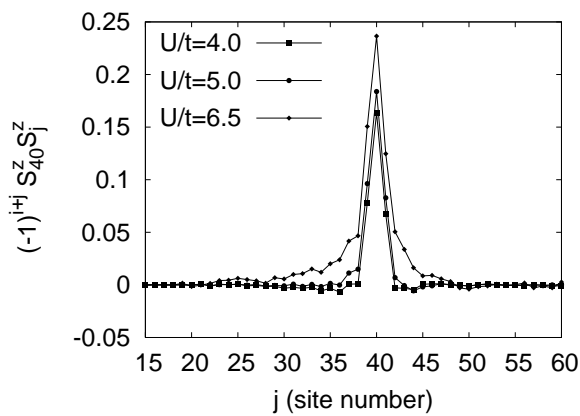}
\end{center}
\caption{Staggered long-range spin correlations 
$(-1)^{i+j}S_i^z S_j^z$ 
between the 40th site and the $j$th site. 
Squares, circles and diamonds are the results for the system 
with $U=4.0, 5.0$ and $6.5$.}
\label{LRSC}
\end{figure}

The emergence of the Mott insulating state is also
seen clearly in the two-point spin correlation 
$(-1)^{i+j}\langle S_i^z S_j^z\rangle$ 
($i=40$ fixed) shown in Fig. \ref{LRSC}. 
At $U=4$, the local particle density, $\langle n_i \rangle \sim 1.2$,
 at the $40$th site is larger than half filling, which means
that the metallic state is realized around there, and therefore
the spin correlations are extended up to a few neighbor sites.
On the other hand, in the large $U$ case, we find that
the spin correlation length becomes much longer, signaling
 the formation of  the Mott insulating state when $U>6$.

Summarizing, by including on-site and intersite particle 
correlations, our variational theory 
properly reproduces the coexistence phase, which is consistent with 
those of QMC and DMRG simulations.\cite{Rigol, RigolPRA, Yamashita}
Furthermore, by comparing the results obtained from the two trial
 states eqs. (\ref{gwf}) and (\ref{sDHwf}), we can say that 
intersite DH correlations  
are essential to describe the coexistence of metallic and 
Mott insulating states in the confined system.

\subsection{Further improvements of the paramagnetic state}\label{3.3}
Here, we wish to briefly comment on further possible improvements of the
trial wave function for the paramagnetic state.
We examine a trial state which incorporates
the effect of intersite spin correlations.
The corresponding trial state we consider is
\begin{eqnarray}
|\Psi_{S}\rangle &=& \mathcal{P}_{S} |\Psi_{G}\rangle, \label{PSwf} \\
|\Psi_{SQ}\rangle &=& \mathcal{P}_{S} |\Psi_{Q}\rangle. \label{PSQwf} \\
\mathcal{P}_{S} &=&
\exp \left[ \sum_{i} \upsilon_{i} \widehat{\mathcal{S}}_{i} \right] \label{PS}
\end{eqnarray}
where the spin correlation operator 
$\widehat{\mathcal{S}}_{i}$ is defined as,
\begin{equation}
\widehat{\mathcal{S}}_{i} = 
\frac{1}{2} \left( S^{z}_{i} S^{z}_{i-1} + S^{z}_{i} S^{z}_{i+1} \right).
\label{S}
\end{equation}
The ground state energy calculated for each state is 
shown in Fig. {\ref{Egraph}}.
As seen in the figure, we cannot find a drastic change 
in the energy (as well as the physical quantities) 
although a lot of variational parameters for
spin correlations are introduced.
The results indicate that the nearest-neighbor 
$S^z$-$S^z$ spin correlations are not so
important in comparison with the DH correlations.
We think that $\mathbf{S}_{i} \cdot \mathbf{S}_{i+1}$ spin correlations
or longer-range spin correlations are necessary to improve the trial state.

One may also take into account {\it site-dependent} DH correlations
to improve the trial state as,
\begin{equation}
|\Psi_{Q'}\rangle = \exp \left[ \sum_{j} \alpha'_{i} 
\widehat{Q}_{i} \right]|\Psi_G\rangle.
\label{mDHwf}
\end{equation}
In general, it is difficult to find the ground-state parameters
 for this wave function in the strong coupling
regime even by means of the SRH scheme. 
Nevertheless, by employing smaller lattice 
systems, we have checked  
that the site-dependent DH correlations give similar
results to that of the {\it uniform} DH correlations.  
Further improvements of the iterative technique
should allow us to discuss the effects of the site-dependent DH
correlations more precisely, which is now under consideration.

\subsection{Examination of the magnetically ordered state}

We discuss here the ground state properties 
by exploiting a trial state including the magnetic order.
It is known that the Hubbard model in the periodic lattice without
confining potentials
 has a tendency to stabilize the long-range magnetic order 
when the particle density satisfies the 
commensurability, {\it i.e.} the condition of half filling.
In the 1D Hubbard model, however, the real long-range order 
is not realized \cite{Lieb,Takahashi,Shiba,Usuki} due to 
 large quantum fluctuations, in contrast to the 2D 
 square-lattice system
where the magnetic order is stabilized.\cite{Hirsch} 
Nevertheless, it is expected that the effect of 
enhanced spin correlations can be incorporated to
some extent
even in terms of a Hartree-Fock type approximation.

In our inhomogeneous system, 
the spatially-modulated magnetic properties may show up.
Therefore, to examine a trial state with magnetic order, 
we start with the unrestricted 
Hartree-Fock (UHF) state, where we determine
the spatially modulated particle density and 
magnetization at each site self-consistently.\cite{UHF}

\begin{figure}[htb]
\begin{center}
\includegraphics[width=8cm]{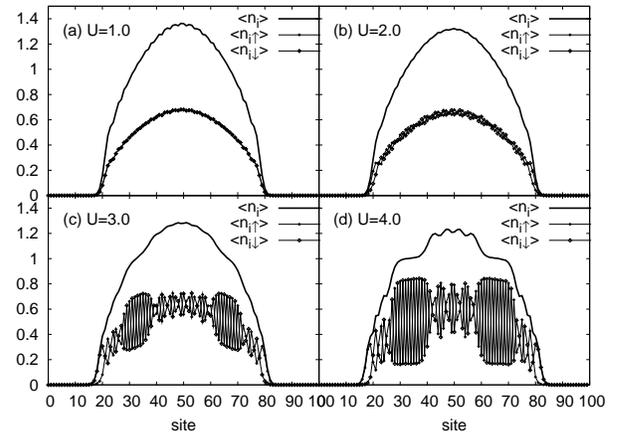}
\end{center}
\caption{
Filled (open) circles represent the density distributions of particles with
 $\uparrow$ ($\downarrow$) spin and a solid line the local particle
 density,
when (a) $U=1.0$, (b) $U=2.0$, (c) $U=3.0$, and (d) $U=4.0$.}
\label{AFseries}
\end{figure}

\begin{figure}[htb]
\begin{center}
\includegraphics[width=8cm]{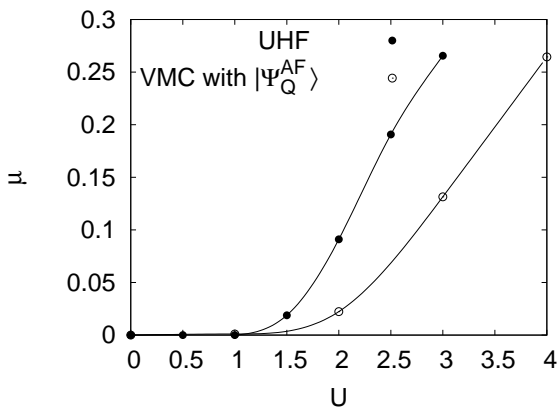}
\end{center}
\caption{The average of the staggered magnetization $\mu$ as a function of $U$.
Filled and open circles represent the magnetization 
obtained by the UHF approximation and the VMC method with 
the trial state $|\Psi_Q^{AF} \rangle$.}
\label{muGraph}
\end{figure}

The trial state for VMC simulations is then written as
\begin{equation}
| \Psi_{Q}^{AF} \rangle = \mathcal{P}_{Q} 
\mathcal{P}_{G} | \Phi_{AF} (M)\rangle,
\label{PQAFwf}
\end{equation}
where $| \Phi_{AF} (M)\rangle$ is the non-interacting state
obtained by the UHF approximation
with an additional variational parameter $M$,
which corresponds to the total magnetization.
Note that the SRH algorithm can be applied only to the Jastrow-type
variational parameters.
Therefore, by performing the SRH scheme with a fixed $M$,
we optimize the variational parameters $(\{\alpha_i\}, \alpha')$.
We show the particle density $\langle n_i \rangle$ and the average of 
the staggered magnetization $\mu$
in Figs. {\ref{AFseries}} and {\ref{muGraph}}.
The latter quantity is defined as,
\begin{equation}
\mu =
\frac{1}{L} \sum_{i}
|\langle n_{i \uparrow} \rangle - \langle n_{i \uparrow} \rangle |.
\label{mu}
\end{equation}
%
%
We find that the magnetization is almost zero and 
little difference appears in the spin-dependent 
density distributions when $U<2$.
This suggests that the paramagnetic ground state is 
realized in the case. 
On the other hand, the increase of the interaction
yields the spatially-modulated magnetization, 
as shown in Fig. {\ref{AFseries}}.
An important point is that 
the large staggered magnetization is induced
in the vicinity of the $i=30$th and $i=70$th sites,
where
the plateau appears in the curve of the particle density,
as shown in Fig. \ref{AFseries}.
This implies that  
magnetic correlations are strongly enhanced in the {\it finite} 
insulating region, which is consistent with 
those discussed in the previous subsection.

Note that the VMC simulations with the HF
approximation cannot describe a paramagnetic quantum liquid state 
correctly in 1D. Nevertheless, 
we can see that the magnetically ordered state is more stable than 
the paramagnetic state in the large $U$ region within
 our VMC analysis (see also Fig. \ref{Egraph}).
Therefore, if we can properly incorporate quantum fluctuations that
 suppress the magnetic order in the Hartree-Fock state, we end up with 
an excellent trial state. 
Alternatively, starting from the paramagnetic state,
we should incorporate spatially extended spin correlations
more precisely. These problems are now under consideration.

\section{Summary and Discussions}

We have investigated correlation effects in
the 1D Hubbard model with harmonic confinement,
which may be relevant to correlated fermionic systems trapped
in the optical lattices.
We have aimed at developing an efficient method based on 
 the VMC simulations that is tractable and 
applicable to the optical lattice systems.
One of the central problems to be resolved is how to treat a large number
of site-dependent variational parameters.

As a first step to incorporate the effect of particle correlations 
in such inhomogeneous systems, we have examined a
trial state with site-dependent Gutzwiller factors (on-site correlations).  
By exploiting the stochastic
reconfiguration with Hessian acceleration scheme, we were able 
to treat more than 50 Gutzwiller variational parameters, whereas
the standard Gutzwiller approach deals with only one or two 
 parameters.
We have thus clarified how the introduction of 
the repulsive interaction changes the distribution of local quantities. 
It turns out that although site-dependent correlations 
are taken into account, the Mott insulating region cannot 
be properly described only in terms of the on-site Gutzwiller factors.

We have thus improved the trial state by taking into account 
the intersite DH correlations. It has been elucidated that the 
coexistence state with  metallic and  Mott insulating regions 
emerges for large $U$ in this case, which is consistent 
with those obtained by other numerical studies.
 This suggests that the 
intersite DH correlations are essential to describe
the metal-insulator coexistence phase.  We have further examined 
how the intersite spin correlations modify the results, 
and have found that as far as Ising-type spin correlations are
concerned, the results are little changed. Nevertheless, there still  
remains a possibility to improve the results by incorporating 
the Heisenberg-type isotropic spin correlations. It has been 
also shown that the variational 
treatment starting from the mean-field antiferromagnetic state
reasonably reproduces the coexistence of metallic and 
insulating regions,
although the real long-range order should be prohibited 
 due to large quantum fluctuations inherent in 1D systems.
In this way, the present scheme is quite flexible, thereby
providing  us with various ways to improve our variational
treatment of the Hubbard model with a site-dependent potential.

In conclusion, we have confirmed that 
the VMC simulations with numbers of site-dependent parameters
can be efficiently performed 
 with the aid of the stochastic acceleration scheme.
This demonstrates that the stochastic VMC method provides
 a powerful tool to treat the
 correlation effects in optical lattice systems with a site-dependent 
potential, which have not been easily tractable by conventional 
optimization procedures.
This naturally motivates us to extend our discussions to the
 2D  and 3D optical lattice systems, 
which is now under consideration.

\section*{Acknowledgment}

The numerical computations were carried out at the Supercomputer Center, 
the Institute for Solid State Physics, University of Tokyo.
We would like to thank M. Yamashita for valuable discussion.
This work was supported by Grant-in-Aids for Scientific Research 
[Grant No. 17740226 (AK) and Grant No. 18043017 (NK)] 
from The Ministry of Education, Culture, Sports, Science 
and Technology of Japan.



\begin{thebibliography}{99}

\bibitem{Greiner} 
M. Greiner, O. Mandel, T. Esslinger, T. W. H{\"a}nsch and I. Bloch:
Nature (London) \textbf{415} (2002) 39.

\bibitem{Nature} 
For a review, see Nature (London) \textbf{416} (2002) 205-246.

\bibitem{Feshbach} 
S. Inouye, M. R. Andrews, J. Stenger, H. -J. Miesner, D. M. Stamper-Kurn and 
W. Ketterle:
Nature (London) \textbf{392} (1998) 151.


\bibitem{Mandel} 
O. Mandel, M. Greiner, A. Widera, T. Rom, T.W. H{\"a}nsch and I, Bloch:
Nature (London) \textbf{425} (2003) 937.

\bibitem{Stoferle2004} 
T. St{\"o}ferle, H. Moritz, C. Schori, M. K{\"o}hl and T. Esslinger:
Phys. Rev. Lett. \textbf{92} (2004) 130403.

\bibitem{Paredes} 
B. Paredes, A. Widera, V. Murg, O. Mandel, S. F{\"o}lling, I. Cirac, G. V. Shlyapnikov, T.W. H{\"a}nsch and I. Bloch:
Nature (London) \textbf{429} (2004) 277.

\bibitem{Gerbier} 
F. Gerbier, A. Widera, S. F{\"o}lling, O. Mandel, T. Gericke and I. Bloch:
Phys. Rev. Lett. \textbf{95} (2005) 050404.

\bibitem{Gerbier2006} 
F. Gerbier, S. F{\"o}lling, A. Widera, O. Mandel, T. Gericke and I. Bloch:
Phys. Rev. Lett. \textbf{96} (2006) 090401.


\bibitem{Jaksch} 
D. Jaksch, C. Bruder, J. I. Cirac, C. W. Gardiner and P. Zoller:
Phys. Rev. Lett. \textbf{81} (1998) 3108.

\bibitem{NTTYamashita} 
M. W. Jack and M. Yamashita:
Phys. Rev. A \textbf{67} (2003) 033605.

\bibitem{Pollet} 
L. Pollet, S. Rombouts, K. Heyde and J. Dukelsky:
Phys. Rev. A \textbf{69} (2004) 043601.

\bibitem{Bergkvist} 
S. Bergkvist, P. Henelius, and A. Rosengren:
Phys. Rev. A \textbf{70} (2004) 053601.

\bibitem{Kovrizhin} 
D. L. Kovrizhin, G. Venketeswara, and S. Sinha:
Europhys. Lett. \textbf{72} (2005) 162.

\bibitem{Scarola} 
V. W. Scarola and S. D. Sarma:
Phys. Rev. Lett. \textbf{95} (2005) 033003.

\bibitem{Wessel} 
S. Wessel, F. Alet, S. Trebst, D. Leumann, M. Troyer and G. G. Batrouni:
J. Phys. Soc. Jpn \textbf{74} (2005) 10.

\bibitem{Kollath} 
C. Kollath, U. Schollw{\"o}ck, J. V. Delft and W. Zwerger:
Phys. Rev. A \textbf{71} (2005) 053606.


\bibitem{DeMarco} 
B. DeMarco and D. S. Jin:
Science \textbf{285} (1999) 1703.

\bibitem{Truscott} 
A. G. Truscott, K. E. Strecker, W. I. McAlexander, G. B. Partridge and
R. G. Hulet:
Science \textbf{291} (2001) 2570.

\bibitem{Schreck} 
F. Schreck, L. Khaykovich, K. L. Corwin, G. Ferrari, T. Bourdel, J.
Cubizolles and C. Salomon:
Phys. Rev. Lett. \textbf{87} (2001) 080403.

\bibitem{Granade} 
S. R. Granade, M. Gehm, K. M. O'Hara and J. E. Thomas:
Phys. Rev. Lett. \textbf{88} (2002) 120405.

\bibitem{Hadzibabic} 
Z. Hadzibabic, C. A. Stan, K. Dieckmann, S. Gupta, M. W. Zwierlein, A.
G{\"o}rlits and W. Ketterle:
Phys. Rev. Lett. \textbf{88} (2002) 160401.

\bibitem{Greiner2004} 
M. Greiner, C. A. Regal, and D. S. Jin:
Nature (London) \textbf{429} (2004) 277.

\bibitem{Bartenstein} 
M. Bartenstein, A. Altmeyer, S. Riedl, S. Jochim, C. Chin, J. H. Denschlag and R. Grinmm:
Phys. Rev. Lett. \textbf{92} (2004) 120401.

\bibitem{Zwierlein} 
M. W. Zwierlein, C. A. Stan, C. H. Schunck, S. M. F. Raupach, A. J. Kerman and W. Ketterle:
Phys. Rev. Lett. \textbf{92} (2004) 120403.


\bibitem{Roati} 
G. Roati, E. de Mirandes, F. Ferlaino, H. Ott, G. Modugno and M. Inguscio:
Phys. Rev. Lett. \textbf{92} (2004) 230402.

\bibitem{Kohl} 
M. K{\"o}hl, H. Moritz, T. St{\"o}ferle, K. G{\"u}nter and T. Esslinger:
Phys. Rev. Lett. \textbf{94} (2005) 080403.

\bibitem{Stoferle}
T. St{\"o}ferle, H. Moritz, K. G{\"u}nter, M. K{\"o}hl and T. Esslinger:
Phys. Rev. Lett. \textbf{96} (2006) 030401.


\bibitem{Rigol} 
M. Rigol, A. Muramatsu, G. G. Batrouni and R. T. Scalettar:
Phys. Rev. Lett. \textbf{91} (2003) 130403.

\bibitem{RigolPRA} 
M. Rigol and A. Muramatsu:
Phys. Rev. A \textbf{69} (2004) 053612.

\bibitem{Pour} 
F. K. Pour, M. Rigol, S. Wessel and A. Muramatsu:
cond-mat/0608491.

\bibitem{Yamashita} 
T. Yamashita, N. Kawakami and M. Yamashita:
unpublished.

\bibitem{Yamashita2} 
T. Yamashita, N. Kawakami and M. Yamashita:
Phys. Rev. A {\it in press}.


\bibitem{McMillan} 
W. L. McMillan:
Phys. Rev. {\bf 138} (1965) A442 .

\bibitem{Ceperley}
D. Ceperley, G. V. Chester and K. H. Kalos:
Phys. Rev. B {\bf 16} (1977) 3081.

\bibitem{Yokoyama}
H. Yokoyama and H. Shiba:
J. Phys. Soc. Jpn. \textbf{56} (1987) 1490;
J. Phys. Soc. Jpn \textbf{56} (1987) 3582;
J. Phys. Soc. Jpn. \textbf{59} (1990) 3669.

\bibitem{YokoyamaP}
H. Yokoyama:
Prog. Theor. Phys. \textbf{108} (2002) 59.


\bibitem{Sorella}
S. Sorella:
Phys. Rev. B \textbf{71} (2005) 241103.



\bibitem{Giamarchi}
T. Giamarchi and C. Thuillier: 
Phys. Rev. B \textbf{42} (1990) 10641; \textbf{43} (1991) 12943.

\bibitem{Oguri}
A. Oguri and T. Asahata:
Phys. Rev. B \textbf{46} (1992) 14073.

\bibitem{Otsuka}
H. Otsuka:
J. Phys. Soc. Jpn. \textbf{61} (1992) 1645.

\bibitem{Liu}
Y. Liu, J. Dong, C. Gong, and T. Chen:
Phys. Rev. B \textbf{48} (1993) 1306.

\bibitem{WatanabeII}
T. Watanabe, H. Yokoyama, Y. Tanaka and J. Inoue:
J. Phys. Soc. Jpn. \textbf{75} (2006) 074707.

\bibitem{Kuratani}
S. Kuratani, A. Koga, and N. Kawakami:
unpublished.


\bibitem{Kobayashi}
K. Kobayashi and H. Yokoyama:
J. Phys. Chem. Solid \textbf{66} (2005) 1384.

\bibitem{Koga}
A. Koga, N. Kawakami, H. Yokoyama, and K. Kobayashi:
AIP Conf. Proc. \textbf{850} (2006) 1458.


\bibitem{WatanabeI}
T. Watanabe, H. Yokoyama, Y. Tanaka, J. Inoue, and M. Ogata:
J. Phys. Soc. Jpn. \textbf{73} (2004) 3404.


\bibitem{McQueen}
P. G. McQueen and C. S. Wang:
Phys. Rev. B \textbf{44} (1991) 10021.


\bibitem{Kono}
H. N. Kono and Y. Kuramoto:
J. Phys. Soc. Jpn. \textbf{75} (2006) 084706.


\bibitem{Jastrow}
R. Jastrow:
Phys. Rev. \textbf{98} (1955) 1479.


\bibitem{Gutzwiller}
M. C. Gutzwiller:
Phys. Rev. Lett. \textbf{10} (1963) 159; 
Phys. Rev. \textbf{134} (1964) A923.

\bibitem{Kaplan}
T. A. Kaplan, P. Horsch and P. Flude:
Phys. Rev. Lett. \textbf{49} (1982) 889.

\bibitem{Fazekas}
P. Fazekas and K. Penc:
Int. J. Mod. Phys. B \textbf{2} (1988) 485.


\bibitem{Lieb}
E. H. Lieb and F. Y. Wu:
Phys. Rev. Lett. \textbf{20} (1968) 1445.

\bibitem{Takahashi}
M. Takahashi:
Prog. Theor. Phys. \textbf{42} (1969) 1098;
\textbf{45} (1971) 756.

\bibitem{Shiba}
H. Shiba:
Phys. Rev. B \textbf{6} (1972) 930.

\bibitem{Usuki}
T. Usuki, N. Kawakami and A. Okiji:
Phys. Lett. \textbf{135A} (1989) 476.


\bibitem{Hirsch}
J. E. Hirsch:
Phys. Rev. Lett. \textbf{51} (1983) 1900; 
Phys. Rev. B \textbf{31} (1985) 4403.



\bibitem{UHF}
N. Furukawa and M. Imada:
J. Phys. Soc. Jpn. \textbf{60} (1991) 3669.






\end{thebibliography}
\end{document}